\documentclass[floatfix,showpacs,amsmath,amsfonts,amssymb,aps,twocolumn,10pt]{revtex4-1}

\usepackage{color,graphicx}
\usepackage{amsmath,amsfonts,amssymb,url,amsthm}

\newcommand{\ket}[1]{\left| #1 \right\rangle}
\newcommand{\bra}[1]{\left\langle #1 \right|}

\newcommand{\pos}[0]{\mathbf{r}}

\newcommand{\Pos}[0]{\mathbf{R}}

\newcommand{\pdiff}[2]{\frac{\partial #1}{\partial #2}}

\newcommand{\figref}[1]{Fig.\ \ref{#1}}
\newcommand{\appref}[1]{App.\ \ref{#1}}

\newcommand{\kf}[0]{k_\mathrm{F}}
\newcommand{\Ef}[0]{E_\mathrm{F}}
\newcommand{\punc}[1]{\;#1}
\newcommand{\diffd}[0]{\mathrm{d}}
\newcommand{\Emax}[0]{E_\mathrm{max}}
\newcommand{\kmax}[0]{k_\mathrm{max}}

\newcommand{\todo}[1]{}

\begin{document}

\title{High-fidelity pseudopotentials for the contact interaction}
\author{P.O.~Bugnion}
\author{P.~L\'opez R\'ios}
\author{R.J.~Needs}
\author{G.J.~Conduit}
\affiliation{Cavendish~Laboratory, J.J.~Thomson~Avenue, Cambridge, CB3~0HE, United~Kingdom}
\date{\today}

\begin{abstract}
  The contact interaction is often used in modeling ultracold atomic gases,
  although it leads to pathological behavior arising from the divergence of the
  many-body wavefunction when two particles coalesce.  This makes it
  difficult to use this model interaction in quantum Monte Carlo and other
  popular numerical methods. Researchers therefore model the contact
  interaction with pseudopotentials, such as the square well
  potential, whose scattering properties deviate markedly from those of the
  contact potential.  In this article, we propose a family of
  pseudopotentials that reproduce the scattering phase shifts of the contact
  interaction up to a hundred times more accurately than the square well
  potential. Moreover, the
  pseudopotentials are smooth, resulting in significant improvements in
  efficiency when used in numerical calculations.
\end{abstract}


\maketitle

Interactions between particles are central to our understanding of
correlated phenomena.  The contact potential is often used to model
interactions in ultracold atomic gases but, despite its widespread use, it
displays pathological behavior: both the potential and wavefunction diverge
when two particles coalesce.  These divergences impede numerical methods,
and are commonly handled by replacing the contact potential by a
pseudopotential, such as a hard sphere or a square well potential.  However,
these approximations to the contact potential display incorrect variations
in the scattering phase shift with incident particle
energy~\cite{Astrakharchik04i,Astrakharchik04ii,Conduit09,Pilati10,
  Chang10,Giorgini99}.  In this article, we adapt methods commonly used for
the development of electron-ion pseudopotentials in the electronic structure
community to propose a new atom-atom pseudopotential whose scattering
properties agree closely with those of the contact interaction.

Ultracold atomic gases have delivered many important insights into strongly 
correlated systems. They can provide both clean model Hamiltonians and 
introduce the ability to tune the strength of the contact interactions.
Ultracold atoms interact through an underlying attractive van der
Waals interaction. An external magnetic field can be used to  
tune the energy of the bound
molecular state to approach the energy of the scattering state, causing the
states 
to couple resonantly. The effect of the resonance on the scattering state
can be modelled by an effective interparticle potential~\cite{Chin09}. In the 
case
of broad Feshbach resonances, the scattering can be described by a single 
universal
scattering length $a$ which describes the scattering phase shift arising from 
a
contact interaction. There are three types of contact interaction:
sufficiently deep to trap a two-body bound state ($a>0$), weakly attractive
with no bound state ($a<0$), and repulsive (excited state of the $a>0$
potential). 

Contemporary numerical simulations of the first two types, the bound state
($a>0$) and weak attractive interactions ($a<0)$, normally adopt a finite
ranged square well or P\"oschl-Teller interaction.  Such simulations have
delivered crucial insights into Bose gases~\cite{Astrakharchik04ii} and the
crossover from a gas of weakly coupled Bardeen-Cooper-Schrieffer pairs
to a strongly-interacting Bose-Einstein
condensate~\cite{Astrakharchik04i,Morris10}, as well as few-atom
physics~\cite{Bugnion13,Bugnion13i,Bugnion13ii}. However, the finite range
imbues the potential with incorrect scattering properties. Reducing the
range of the potential alleviates this problem, but slows numerical
calculations.

The third type of contact potential gives repulsive interactions that drive
itinerant ferromagnetism in Fermi gases~\cite{Conduit09i,Pilati10,Chang10},
a Tonks-Girardeau gas~\cite{Astrakharchik04i}, and a Bose
gas~\cite{Giorgini99}. The repulsive interaction emerges from the first
excited state of the bound state potential. Both the repulsive contact
potential and
the bound state potential therefore have $a>0$. In ultracold atomic gas
experiments~\cite{Jo09} the excited state (also called the upper branch) is
protected from decay to the ground state by a slow three-body loss process,
allowing the study of repulsive interactions. To simulate these repulsive
interactions, one can adopt a finite-ranged attractive
potential and study the first excited eigenstate~\cite{Pilati10,Chang10}.
However, studying excited states in quantum Monte Carlo (QMC) methods often
requires restricting the excited state wavefunction to be orthogonal to the
lower energy states.  Variational estimates of excited state energies
calculated within the widely-used diffusion quantum Monte Carlo (DMC) method
\cite{Ceperley80,Umrigar93,Foulkes01} are discussed in Ref.\
\cite{Symmetry_Constraints_DMC}. The fixed node constraint used in DMC
prevents collapse into the ground state, but it is still difficult to
calculate reliable excited state energies within DMC. An alternative
approach is to use a repulsive top-hat potential~\cite{Conduit09i} whose
ground state resembles the first excited state of the contact
potential. However, this potential has a finite range greater than the
scattering length, resulting in an incorrect scattering phase shift.

The difficulty in simulating repulsive interactions means that there are
important open questions about fermionic gases: is the ground state of a
strongly interacting fermionic system
ferromagnetic~\cite{Conduit08,Conduit09i,Conduit09,Zhai09,Duine05,
  Maslov09,Chubukov09,Pedder13}; is the ferromagnetic transition first or
second order; and whether exotic phases emerge around quantum criticality
such as spin spirals~\cite{Conduit09i}, nematic
phases~\cite{Chubukov09,Chubukov10,Karahasanovic12}, and a counterintuitive
$p$-wave superconducting
state~\cite{Balian63,Fay80,Mathur98,Roussev01,Conduit13,Saxena00,Huxley01}. The
development of a pseudopotential that is better able to reproduce the
scattering properties of the contact interaction will help resolve these
open questions.

In section I, we present two pseudopotentials for the interatomic
interaction in a cold atom gas.  We first adapt norm-conserving
pseudopotentials~\cite{Hamann79,Bachelet82,Hamann89}, developed by the
electronic structure community for electron-ion interactions, to deal
effectively with scattering states. We then present a new pseudopotential
constructed to minimize the scattering phase shift error for all wavevectors
in a Fermi gas.  In section II, we test the accuracy of the new formalism
using the exactly soluble system of two trapped atoms. In section III, we
investigate how the pseudopotential performs in a many-body setting by
calculating the equation of state of the weakly repulsive Fermi gas and
comparing results to a perturbation expansion.

\section{Derivation of the pseudopotentials}

\begin{figure}
 \includegraphics[width=\linewidth]{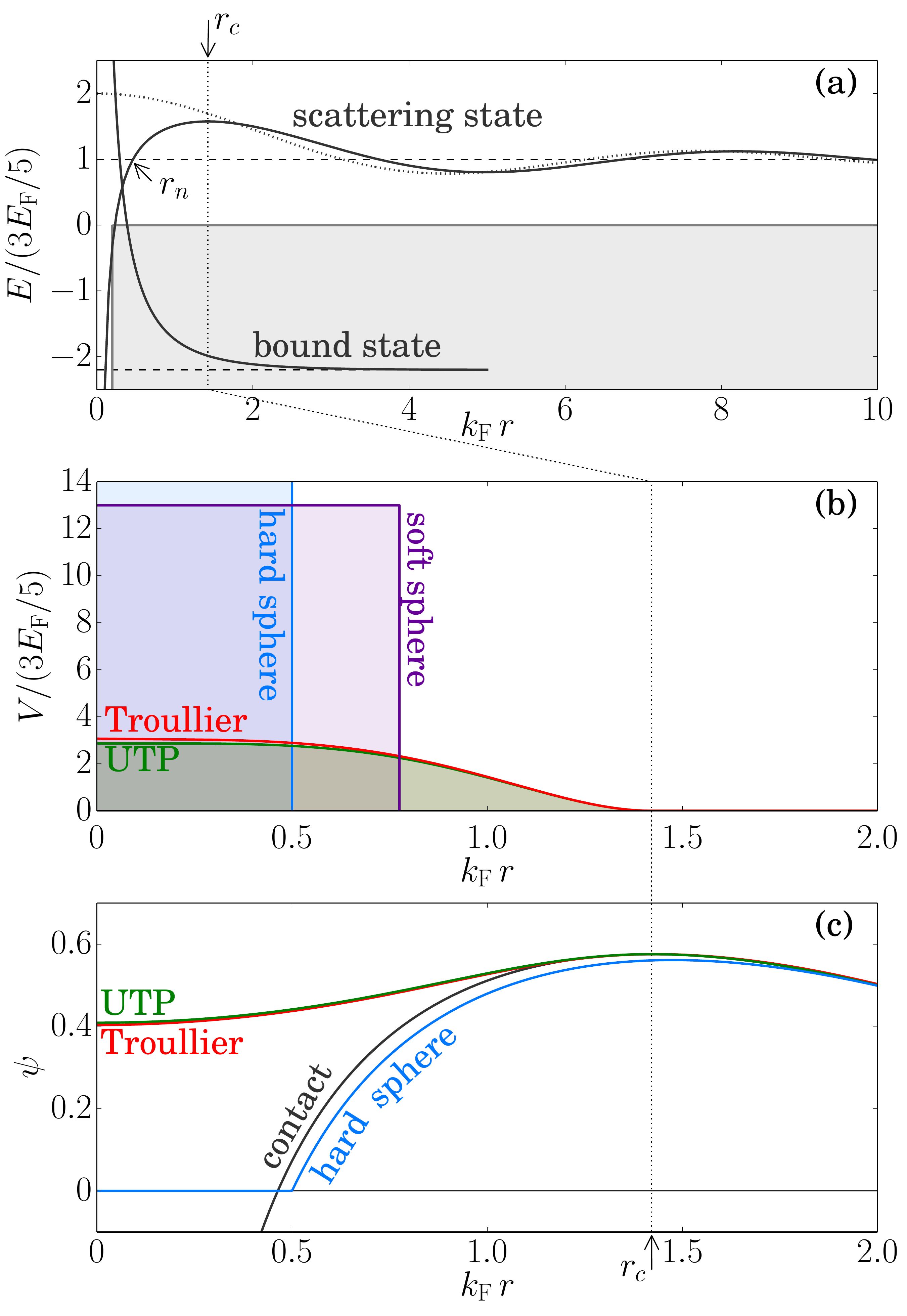}
 \caption{(Color online) (a) Bound state and scattering wavefunctions for
   contact interactions at $\kf a = 0.5$, offset by their respective
   eigenvalues (dashed lines) as a function of inter-particle
   separation. The bare contact potential (represented by the gray area) is
   strongly attractive and harbors a single bound state.  The scattering
   states incident on the potential incur a positive phase shift with
   respect to the non-interacting scattering wavefunction (dotted
   line). $r_{\text{n}}$ denotes the first node of the scattering
   wavefunction and $r_{\text{c}}$ denotes the first antinode, which we use
   as the cutoff radius when constructing pseudopotentials, as described in
   section~\ref{subsec:repulsive}.  (b) The pseudopotentials at $\kf
   a\!=\!1/2$ on the repulsive branch.  The potential labeled ``Troullier''
   denotes the pseudopotential derived using the Troullier-Martins
   formalism. The line labeled ``UTP'' denotes the pseudopotential derived
   using the UTP formalism. These formalisms are described in
   section I.  (c) The wavefunctions for the relative motion of two
   particles interacting with a contact potential, the hard sphere,
   Troullier-Martins pseudopotential and UTP, at $k=\kf$.}
 \label{fig:potentials}
\end{figure}

\begin{figure}
 \includegraphics[width=1.0\linewidth,trim=0 0 0 0,clip=true]{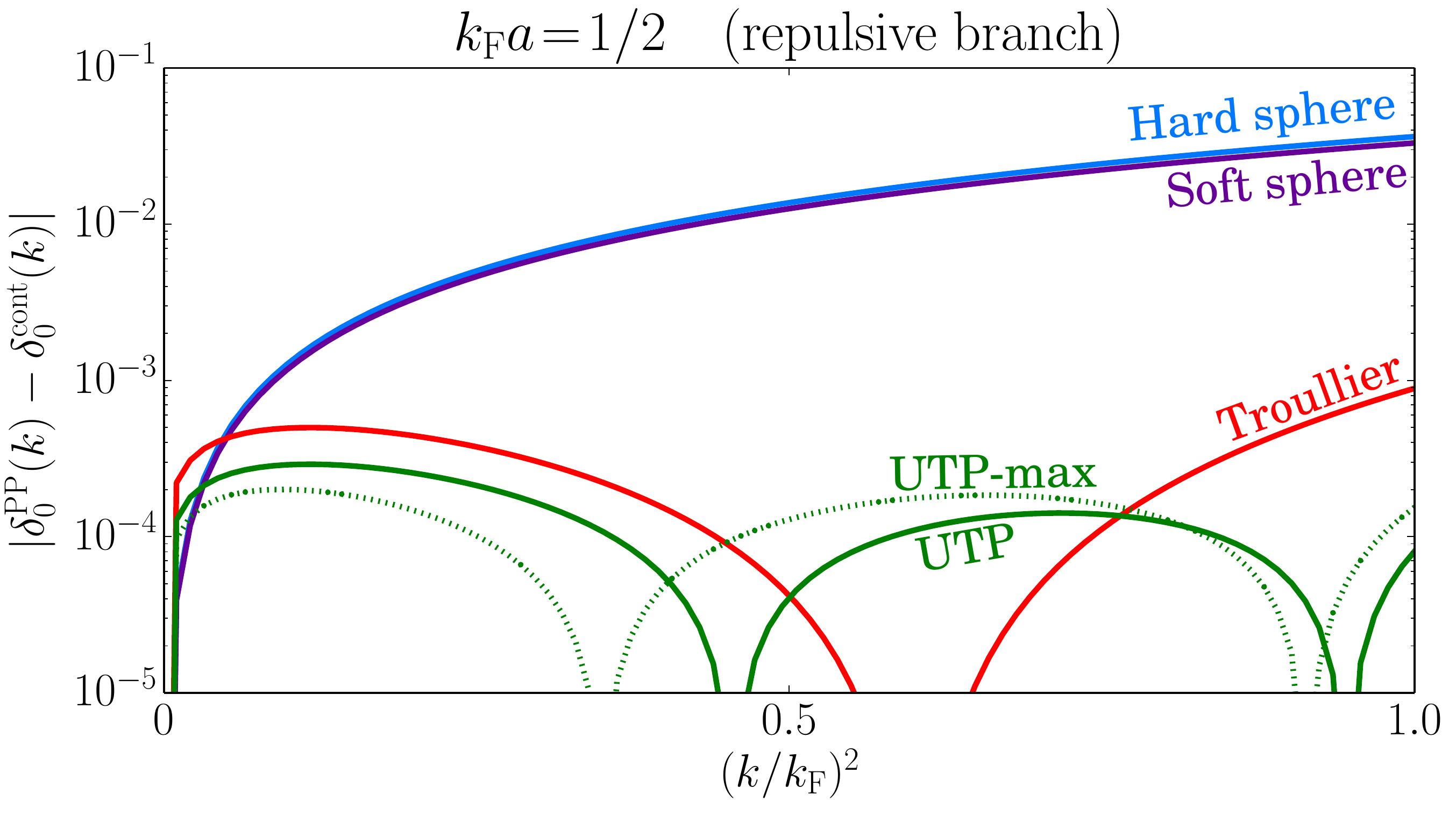}
 \caption{(Color online) The errors in phase shifts
   $|\delta_0^\mathrm{PP}(k) - \delta_0^\mathrm{cont}(k)|$ for the repulsive
   branch at $\kf a=1/2$. $\delta_0^\mathrm{cont}$ is the $s$-wave
   scattering phase shift for the contact interaction and
   $\delta_0^\mathrm{PP}$ is the scattering phase shift for each of the
   pseudopotentials.  The Troullier-Martins formalism is approximately two
   orders of magnitude more accurate than the hard and soft sphere
   pseudopotentials commonly used as approximations to the contact
   interaction. The UTP formalism offers an additional factor of two
   improvement. The dotted green line labeled UTP-max denotes the phase shift 
   error of a variant of the UTP developed by minimizing the peak phase shift error.}
 \label{fig:phase_shifts}
\end{figure}

To construct the pseudopotential we study the two-body problem: two identical fermions
in their center-of-mass frame with wavevector $k\ge0$. The Hamiltonian in
atomic units ($\hbar=m=1$) in the center of mass frame is
\begin{equation*}
-\frac{\nabla^2}{2\mu}\psi+V(\pos)\psi=\frac{k^2}{2\mu}\psi \punc{,}
\end{equation*}
where $V(\pos) = 4\pi a \delta(\pos) (\partial/\partial r)r$ is the contact
potential for scattering length $a$ and inter-particle separation
$\pos$~\cite{Busch98}, and $\mu = 1/2$ is the reduced mass.

The scattering states for the contact potential are
$\psi^{\mathrm{cont}}_{k,\ell}=\sin[kr-\ell\pi/2+\delta^\mathrm{cont}_\ell(k)]/kr$,
where
\begin{equation*}
    \delta^\mathrm{cont}_\ell(k) = \left\{ \begin{array}{ll}
            \arctan(-ka) & \ell = 0 \\
            0 & \ell > 0
        \end{array}  \right. \; \punc{,}
\end{equation*}
is the scattering phase shift in the angular momentum channel $\ell$. We
seek a pseudopotential that:
\begin{enumerate}
\item reproduces the correct phase shifts over the range of wavevectors
  $0\le k\lesssim\kf$ present in a Fermi gas with Fermi wavevector $\kf$,
\item supports no superfluous bound states to be compatible with
  ground state methods,
\item is smooth to accelerate numerical calculations.
\end{enumerate}

We start by developing pseudopotentials for the repulsive branch, then the
attractive branch and finally the bound state. When developing
pseudopotentials, we benchmark their quality by looking at how closely the
phase shift of the wavefunction for the relative motion of two particles
interacting via the pseudopotential reproduces the phase shift of the
contact interaction for all wavevectors $0 \le k \lesssim \kf$ present in a
Fermi gas, as shown in Figs.\ \ref{fig:phase_shifts} and
\ref{fig:delta-cutoff}.

\subsection{Repulsive branch}
\label{subsec:repulsive}

We first focus on developing a pseudopotential for the repulsive branch of
the contact interaction. This branch offers a particular challenge. The bare
potential is strongly attractive, harboring exactly one bound state, as shown
in~\figref{fig:potentials}(a). The excited states of this potential must
maintain orthogonality to the bound state, resulting in a positive phase
shift. The scattering states have one more node than the non-interacting
state with the same wavevector, as shown in~\figref{fig:potentials}(a).

We describe four families of pseudopotentials: hard sphere, soft sphere (top
hat), the Troullier-Martins form of norm-conserving
pseudopotential~\cite{Troullier91,Hamann79} and a pseudopotential that aims
to minimize the error in scattering phase shift over all wavevectors
occupied in the Fermi gas. The first two families (the hard sphere and top
hat) have frequently been used as approximations to the contact potential in
numerical calculations~\cite{Conduit09,Pilati10,
  Chang10,Bugnion13ii,Conduit09i}.

The usual approach~\cite{Pilati10,Conduit09i} to the construction of
pseudopotentials for the contact interaction starts from the low energy
expansion for the $s$-wave scattering phase shift,
\begin{equation}
    \cot\delta_0(k)=-\frac{1}{ka}+\frac{1}{2}kr_\mathrm{eff}-Pr_\mathrm{eff}^3 k
    ^3+\mathcal{O}(k^5)
    \punc{,}
    \label{eq:low-energy-delta}
\end{equation}
where $r_\mathrm{eff}$ is the ``effective range'' of the potential and $P$
is the ``shape parameter''.  For a contact potential, $r_\mathrm{eff}$ and
all higher order terms are zero.  Perhaps the simplest pseudopotential is a
hard sphere potential with radius $a$. This reproduces the correct
scattering length $a$, thereby delivering the correct phase shift for
$k=0$. However, the hard sphere has an effective range
$r_\mathrm{eff}=2a/3$. To study the impact of the pseudopotential on the
scattering states, we calculate the phase shifts at $\kf a=1/2$ for all
wavevectors between 0 and $\kf$ and compare them to the contact phase
shifts.  \figref{fig:phase_shifts} shows that the finite effective range of
the hard sphere potential causes significant deviations in the scattering phase
shift for $k>0$.

To reduce the error in the scattering phase shift, Ref.~\cite{Conduit09i}
adopted a soft sphere potential: $V(r)=V_0\Theta(r-R)$, with $V_0$ and $R$
chosen to reproduce the contact scattering length
$a=R(1-\tanh\gamma/\gamma)$ and effective range
$r_\mathrm{eff}=R[1+\frac{3\tanh\gamma-\gamma(3+\gamma^2)}
{3\gamma(\gamma-\tanh\gamma)^2}]=0$, where $\gamma=R\sqrt{2\mu V_0}$. The
first two terms in the low energy expansion of the phase shift are now
correct, leading to a small reduction in phase shift error as shown in
\figref{fig:phase_shifts}.

The two potentials considered so far display incorrect behavior at large
wavevectors due to the focus on reproducing the correct $k=0$ scattering
behavior. To improve the accuracy we turn to the
Troullier-Martins~\cite{Troullier91} formalism developed for constructing
attractive electron-ion
pseudopotentials~\cite{Hamann79,Zunger79,Bachelet82,Hamann89,Rappe90,Lin93}.
These pseudopotentials reproduce both the correct phase shift and its
derivative with respect to energy at a prescribed calibration energy.  The
Troullier-Martins form of norm-conserving pseudopotential can readily be
applied to the construction of a pseudopotential for the contact
interaction.  We choose a calibration energy and cutoff radius:

\emph{Calibration energy}:\ the pseudopotential will have scattering
properties identical to the contact potential at the calibration energy. For
electron-ion pseudopotentials, the bound state energy in an isolated ion is
a natural choice. For example, in a homogeneous fermionic gas the scattering of
states with incident momenta less than $\sim\kf$ is particularly important,
and therefore we choose a calibration energy equal to the median energy of
the occupied states, $(3/5)\Ef$.

\emph{Cutoff radius}:\ the Troullier-Martins pseudo-wavefunction is
identical to the contact wavefunction outside of the cutoff radius, but has
no nodes inside the cutoff radius, as shown
in~\figref{fig:potentials}(c). We can therefore choose the cutoff radius to
eliminate the bound state: by selecting a radius
$r_{\text{c}}\!>\!r_{\text{n}}$, where $r_{\text{n}}$ is the position of the
first node in the wavefunction, we construct a pseudopotential that does not
have a bound state, as shown in~\figref{fig:potentials}(b).  We choose the
first antinode of the wavefunction at the calibration energy as the cutoff
radius for the pseudopotential.

Having chosen a suitable calibration energy and cutoff radius, we construct
the pseudo-wavefunction. The contact potential exhibits a non-zero phase
shift only when the particles are incident with angular momentum quantum
number $\ell=0$. We therefore concentrate on reproducing the correct
$\ell=0$ behavior in this section. We demonstrate how to eliminate
scattering in higher angular momentum channels in section \ref{subsec:nlpp}.

The functional form of the pseudo-wavefunction in the $\ell=0$ channel at
the calibration energy is
\begin{equation*}
  \psi^\mathrm{PP}(\pos)=\left\{ \begin{array}{ll} \exp(\sum_{i=0}^6 c_i 
      r^{2i})\,\mathrm{Y}_0(\theta,\phi) & r < r_{\text{c}} \\
      \psi^\mathrm{cont}_{k,\ell=0}(r) & r \ge r_{\text{c}}
        \end{array} \right. \punc{,}
\end{equation*}
where $k\!=\!\!\sqrt{(3/5)\Ef}$ is the wavevector at the calibration energy
and $\pos=(r,\theta,\phi)$ is the relative position of the interacting
particles. The coefficients $c_i$ are calculated by demanding continuity of
the pseudo-wavefunction and its first four derivatives at the cutoff radius,
and requiring that the derivative of the phase shift with respect to energy,
$\partial(\cot\delta)/\partial E|_{(3/5)\Ef}$ be the same as that of the
contact interaction at the calibration energy. This last condition, called
the norm-conservation condition, is equivalent to demanding that the total
density enclosed by $r<r_{\text{c}}$ for the pseudo-wavefunction matches
that of the contact wavefunction,
\begin{equation*}
\int_{|\pos|<r_{\text{c}}}
|\psi^\mathrm{PP}(\pos)|^2 \, \diffd \pos = \int_{|\pos|<r_{\text{c}}}
|\psi^\mathrm{cont}_{k,\ell=0}(\pos)|^2 \,  \diffd \pos \punc{.}
\end{equation*}
Finally, we demand that $c_2^2=-5c_4$, to guarantee that the pseudopotential
has zero curvature at the origin. Having constructed the
pseudo-wavefunction at the calibration energy, we invert the Schr\"odinger
equation to obtain the pseudopotential $V^\mathrm{PP}(r)$.  The formalism
for the contact interaction is detailed in
appendix~\appref{app:TroullierMartins}.  We also provide a computer program
to generate the pseudopotential~\cite{supplemental-code}.

By calibrating the pseudopotential at the median incident scattering energy
$E=(3/5)\Ef$, we reduce the error in the scattering phase shift over a broad
range of wavevectors. This generates the pseudopotential shown in
\figref{fig:potentials}(b), whose smoothness leads to improved numerical
stability and efficiency.  \figref{fig:phase_shifts} demonstrates that this
potential is exact at the calibration energy $E=(3/5)\Ef$ and delivers a
hundred-fold decrease in phase shift error across all wavevectors, compared
to the soft sphere pseudopotential.


The Troullier-Martins formalism yields a pseudopotential that reproduces the
contact behavior exactly at the calibration energy, but deviates at other
energies. 
One approach for reducing the deviation is to ensure that higher order
derivatives of the phase shift with respect to energy are equal to those of
the contact interaction. A second option is to impose accurate scattering
at multiple energies through additional parameters.
%
%
%
Here we pursue the natural conclusion
of these approaches by constructing a pseudopotential that minimizes the
error in the phase shifts over all the wavevectors occupied in a Fermi
gas. We derive this pseudopotential below, referring to it as an
``ultratransferable pseudopotential'' (UTP).

The UTP is identical to the contact potential outside of a cutoff radius
$r_{\text{c}}$, but has a polynomial form inside the cutoff,
\begin{align*}
 \frac{V(r)}{\Ef}\!=\!
  \begin{cases}\! \left(1\!-\!\frac{r}{r_{\text{c}}}\right)^{2}\!
\left[v_{1}\left(\frac{1}{2}\!+\!\frac{r}{r_{\text{c}}}\right)\!+\!
\displaystyle\sum_{i=2}^{N_{\text{v}}}v_{i}\left(\frac{r}{r_{\text{c}}}\right)^{i}\right]
&\!\!\!r\le r_{\text{c}}\\
0&\!\!\!r>r_{\text{c}}
  \end{cases}  
\!\!\punc{,}
\end{align*}
with $N_{\text{v}}=9$. The term $(1-r/r_{\text{c}})^{2}$ ensures that
the potential goes smoothly to zero at $r=r_{\text{c}}$ and the term
$v_{1}(1/2+r/r_{\text{c}})$ constrains the potential to have zero gradient at
$r=0$ to allow the pseudo-wavefunction to be as smooth as possible.  
This is advantageous in
quantum chemistry methods in which the absence of a cusp improves 
convergence with respect to basis set size. As with the Troullier-Martins pseudopotential, we choose a
cutoff radius that corresponds to the first antinode of the true
wavefunction, removing the node at $r=r_{\text{n}}$ and therefore
eliminating the bound state.  To calculate the coefficients $\{ v_i \}$, we
minimize the total squared error in the phase shift over all wavevectors 
between 0
and $\kf$,
\begin{equation*}
  \langle(\delta_\ell^\mathrm{PP}-\delta_\ell^\mathrm{cont})^2\rangle=
  \frac{\int_{0}^{\kf}\!\!\left[\delta_\ell^\mathrm{PP}(k)-
      \delta_\ell^\mathrm{cont}(k)
    \right]^{2}w(k)\;\diffd k}{\int_0^{\kf}w(k)\;\diffd k}\punc{,}
\end{equation*}
where the phase shift $\delta_\ell^\mathrm{PP}(k)$ is determined from a
numerical calculation of the scattering solution of the pseudopotential and 
$w(k)=k^2$ is a positive weighting function. We
include a computer program to generate the UTP in the supplemental
material~\cite{supplemental-code}. The computer program starts with 
coefficients
determined from the Troullier-Martins pseudopotential, but we verified that the 
optimization was not stuck in a local minimum by repeating the process with 
different initial coefficients.

As demonstrated in \figref{fig:phase_shifts}, this potential gives an error
in $\delta_0$ of less than $10^{-3}$ for all wavevectors $0\le k\le \kf$
found in a Fermi gas, corresponding to an improvement of two orders of
magnitude over previously used pseudopotentials, and an approximate two-fold
improvement over the Troullier-Martins pseudopotential.

We test the robustness of the UTP construction by generating two additional
variants of the formalism. The first, inspired by the Troullier-Martins
pseudopotential, contains only even terms in the polynomial functional form
of the potential.  For the second variant, rather than minimizing the total
squared phase shift error
$\int_{0}^{\kf}[\delta_\ell^\mathrm{PP}(k)-\delta_\ell^\mathrm{cont}(k)]^2
w(k) {\text{d}}k$, we instead minimize the maximum phase shift error
$\max_{0\le
  k\le\kf}(|\delta_\ell^\mathrm{PP}(k)-\delta_\ell^\mathrm{cont}(k)|)$.
Including only even terms in the polynomial functional form of the
pseudopotential delivers a $1\%$ poorer quality pseudopotential for the same
number of variational parameters. Minimizing the maximum phase shift error
leads to a similar pseudopotential, with a slightly smaller peak error, but
the phase shifts deviate more from those of the contact interaction
elsewhere.  Ultimately the selection of the optimization strategy depends on
the physics of the system: for density waves one should minimize the error
around $k=0$, while for $s$-wave superconductivity one should minimize the
error around $k=k_{\text{F}}$. However, having verified that different
optimization procedures lead to similar high quality pseudopotentials, we
continue with the optimization of the total squared phase shift error.

\subsection{Attractive branch}

We can use a similar procedure to derive Troullier-Martins and
ultratransferable pseudopotentials for the attractive branch, $a<0$. The
main difference from the repulsive branch lies in the choice of cutoff: for
the repulsive branch, the cutoff must lie beyond the first node of the
wavefunction, while for the attractive branch there is no lower bound on
the cutoff.

The smaller the cutoff, the closer the scattering properties of the
pseudopotential approach those of the contact potential. However, reducing the
cutoff comes at the cost of computational efficiency.  For example, in
quantum Monte Carlo simulations, the sampling efficiency of a potential is
proportional to the fraction of configuration space volume in which the
potential is finite, $r_{\text{c}}^3/\Omega$, where $\Omega$ is the
simulation cell volume.

\begin{figure}
 \includegraphics[width=\linewidth]{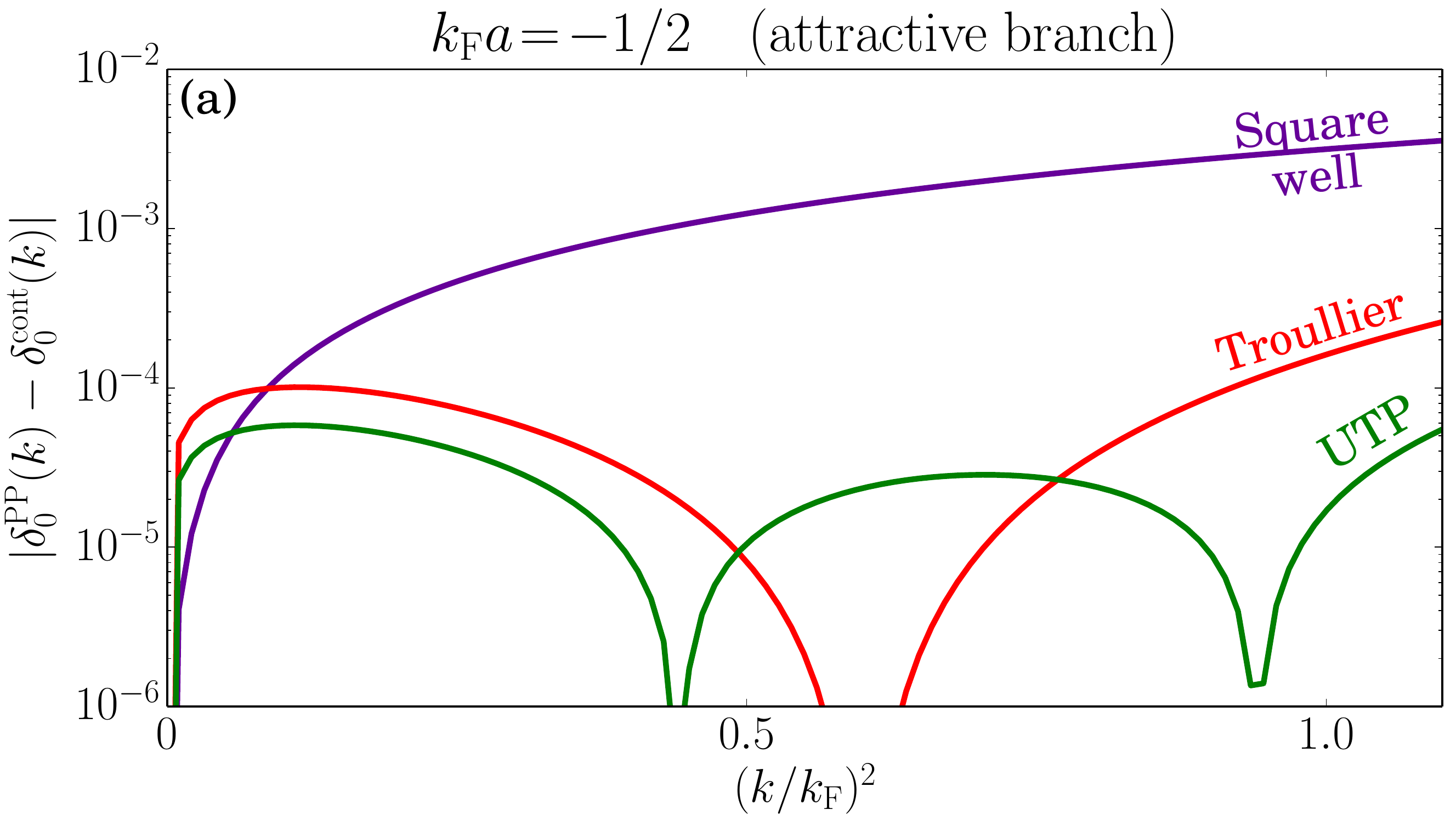}
 \includegraphics[width=\linewidth]{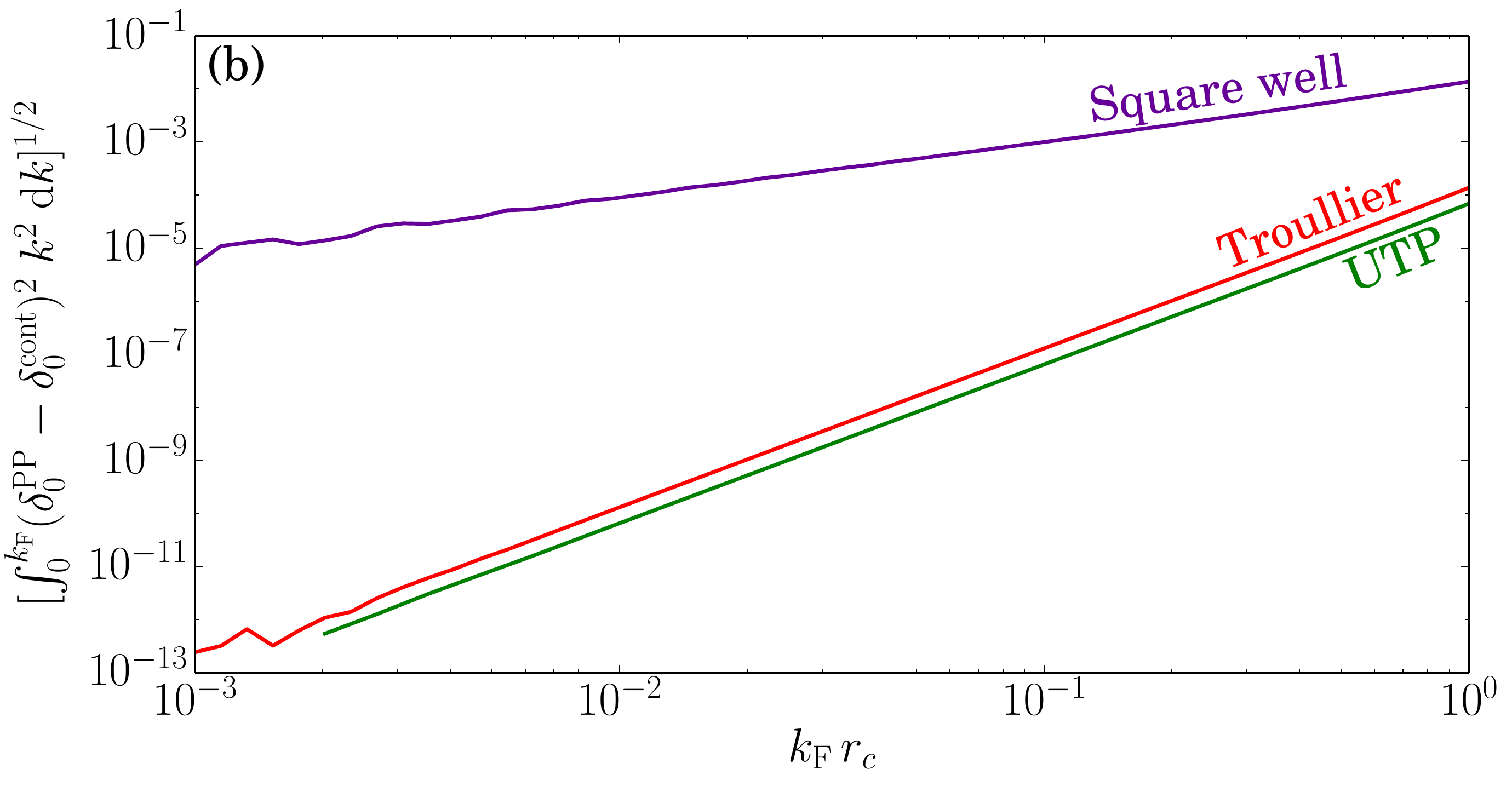}
 \caption{(Color online) (a) The errors in phase shift for the attractive
   branch at $\kf a\!=\!-1/2$, for different pseudopotentials. The
   Troullier-Martins formalism yields phase shifts that are ten times closer
   to those of the contact interaction than the square well
   approximation. For all pseudopotentials described here, the quality of
   the potential depends on the choice of spatial cutoff.  The
   Troullier-Martins and UTP were constructed with cutoff
   $r_{\text{c}}\!=\!(1/2)\kf$. By contrast, the square well potential was
   constructed with $r_{\text{c}}\!\simeq\!0.03\kf$.  (b) Convergence of the
   phase shifts with decreasing pseudopotential radius for
   $k_\mathrm{F}a\!=\!-1/2$.}
 \label{fig:delta-cutoff}
\end{figure}

In~\figref{fig:delta-cutoff}(a) we adopt a cutoff $r_{\text{c}}=1/2\kf$, and
compare to the square well potential with cutoff
$r_{\text{c}}=0.01\sqrt[3]{3\pi^2}/\kf$ used in
Ref.~\cite{Astrakharchik04i}. Both the Troullier-Martins pseudopotential and
the UTP have an average error approximately 10 times smaller than the square
well potential, but their larger cutoff allows them to be sampled 4000
times more efficiently in QMC. Reducing the cutoff used for the
Troullier-Martins pseudopotential or the UTP would increase their accuracy
further, at the cost of a reduction in sampling efficiency.

In~\figref{fig:delta-cutoff}(b), we compare the phase shift accuracy of the
pseudopotentials as a function of cutoff. Both the Troullier-Martins and
ultratransferable formalisms result in pseudopotentials whose scattering
phase shifts converge to those of the contact interaction considerably faster
than the square well potential. We find that the average error in phase shift
of both the Troullier-Martins pseudopotential and UTP tends to zero as
$r_{\text{c}}^3$. By contrast, the square well converges as
$r_{\text{c}}$. The improved convergence can be understood as a consequence
of imposing norm-conservation, which guarantees the correctness of $\partial
(\cot\delta) / \partial E|_{(3/5)\Ef}$ around the calibration
energy $(3/5)\Ef$. Eq.~\eqref{eq:low-energy-delta} then shows that the 
leading error in
the phase shifts is approximately proportional to $\cot(\delta^\mathrm{PP}) -
\cot(\delta^\mathrm{cont}) \propto r_{\text{c}}^3$ for both the 
Troullier-Martins
and UTP. By contrast, for the square well potential, the 
error is
proportional to the effective range, which, in turn, is proportional to
$r_{\text{c}}$.

\subsection{Bound state}

We now construct a pseudopotential for the bound state (corresponding to
$a>0$). Unlike the repulsive and attractive branches described above, all
particles in the bound state exist as tightly bound dimers, with energy $E
\sim -1/2a^2$ per particle. This situation is analogous to that of a valence
electron orbiting an ionic core. The Troullier-Martins formalism therefore
lends itself well to the construction of a pseudopotential for this branch.
We calibrate the Troullier-Martins pseudopotential at the binding energy
(per particle) $E=-1/2a^2$.  The cutoff is constructed in the same manner as
for the attractive branch, delivering a similar improvement in
efficiency. We note that the UTP form is not advantageous for this branch
since all particles have approximately the same energy.

\subsection{Non-local pseudopotentials}
\label{subsec:nlpp}

The pseudopotentials constructed in the previous sections have finite
scattering amplitude in the $p$-wave and higher angular momentum
channels. The contact potential, by contrast, scatters only in the $s$-wave
channel $\ket{s}$. This problem can be solved by using a non-local
pseudopotential $\hat{V}^\mathrm{NL}=\ket{s}V(r)\bra{s}$, where
$\ket{s}\bra{s}$ serves to project out the $s$-wave component of the
wavefunction for the relative motion of the interacting particles, and $V(r)$
is the Troullier-Martins pseudopotential or UTP constructed to reproduce the
scattering properties of the contact interaction in the $s$-wave
channel~\cite{Phillips59,Kleinman60,Austin62}.

Non-local pseudopotentials have been used effectively in quantum Monte Carlo
calculations for the electron-ion interaction~\cite{Fahy90}. Adapting the
formalism to interparticle pseudopotentials is straightforward. The total
contribution to the local energy from the non-local pseudopotential can be
written as a double sum over particles in each spin channel,
\begin{equation*}
    \frac{\hat{V}^\mathrm{NL} \Psi}{\Psi} = \sum_{i \in \uparrow}
    \sum_{j\in\downarrow} \frac{\hat{V}_{ij}^\mathrm{NL} \Psi}{\Psi}\punc{,}
\end{equation*}
where $\Psi$ is the many-body wavefunction.  To calculate the contribution
$\hat{V}^\mathrm{NL}_{ij} \Psi/\Psi$ that arises from the interaction
between an up-spin particle at $\pos_i$ and a down-spin particle at
$\pos_j$, it is convenient to translate all particle positions by $-\pos_i$,
such that particle $i$ is located at the origin. Then,
\begin{equation*}
  \frac{\hat{V}^\mathrm{NL}_{ij} \Psi}{\Psi}
  =  \frac{1}{4\pi} \, V(r_j) \int \frac{\Psi(\Pos^\uparrow\, ; \, \ldots ,{\bf
      r}_j^{\prime},\ldots)}{\Psi(\Pos^\uparrow\,;\,\ldots, \pos_j,\ldots)} \, \diffd\Omega_{{\bf r}_j^{\prime}}
  \punc{,}
\end{equation*}
%
%
where $r_j=|\pos_j|$, $V(r_j)$ is the value of the pseudopotential at $r_j$,
$\Pos^\uparrow$ denotes the positions of all up-spin particles, and the
integration runs over all solid angles on a sphere or radius $r_j$ centered
at the origin.  We note that, inasmuch as the proposed pseudopotentials are
short-ranged, we need only carry out the spherical integration for a small
number of pairs of atoms: all those with $|\pos_i-\pos_j|<r_{\text{c}}$.

Additional accuracy could be gained by using different projectors for
different energy ranges~\cite{Vanderbilt90,Blochl94}. Non-local
pseudopotentials have been used successfully to describe electron-ion
interactions in numerical calculations. The formalism necessary to implement
the projectors is therefore already in place.

\section{Atoms in a trap}

\begin{figure}
 \includegraphics[width=1.0\linewidth,trim=5 0 0 0,clip=true]{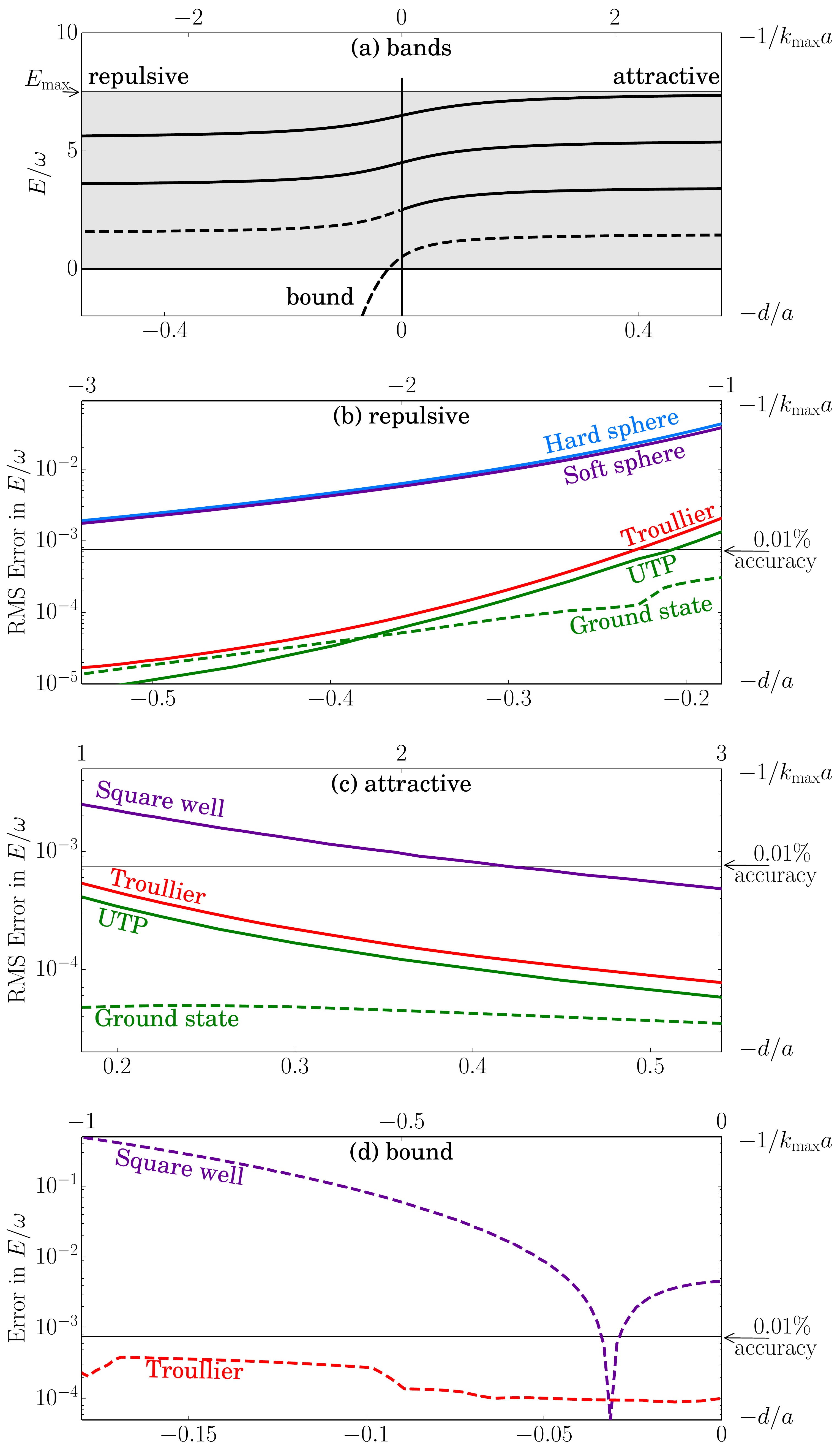}
 \caption{(Color online) (a) Band diagram for two atoms in a harmonic
   trap, calculated following Ref.~\cite{Busch98}.  (b) Mean squared error
   in total energy for two atoms in a harmonic trap, for all bands below
   $\Emax$ (solid lines), for repulsive interactions ($\kmax a>0$). UTP
   denotes the ultratransferable pseudopotential. The dashed line denotes
   the error in the ground state energy with the UTP. The labels $-d/a$ and 
   $-1/k_\mathrm{max}a$ describe the x-axis, which can be
   interpreted as either a change in trap size for constant interaction 
strength
   (varying $d/a$ where $d=1/\sqrt{\omega}$), or a change in interaction 
   strength for constant
   trap size (varying $1/k_\mathrm{max} a$ where 
$k_\mathrm{max}=\sqrt{\Emax}$). The horizontal solid black line
   shows the typical many-body accuracy goal of $0.01\%$. (c) The
   pseudopotential error for attractive interactions ($\kmax a<0$).  (d) The
   pseudopotential error in the bound state energy.}
 \label{fig:busch_tests}
\end{figure}

We have developed a pseudopotential that delivers the correct scattering
phase shift for an isolated system. To test the pseudopotential we turn to
an experimentally realizable configuration~\cite{Serwane11,Zurn12}: two
atoms in a spherical harmonic trap with frequency $\omega$ and
characteristic width $d=1/\sqrt{\omega}$. For all three types of contact
interaction, this system has an analytical solution plotted
in~\figref{fig:busch_tests}(a)~\cite{Busch98} that we can benchmark
against. Moreover, the exact solution extends to excited states, allowing us
to test the performance of the pseudopotential across a wide range of energy
levels to provide a firm foundation from which to study the many-body
system.

\subsection{Ground state}

We first compare the pseudopotential estimates of the ground state energy to
the exact analytical solution~\cite{Busch98}. For the repulsive and
attractive branches, the hard/soft sphere potentials deliver a $\sim1\%$
error in the energy, whilst both the Troullier-Martins and ultratransferable
pseudopotentials (shown in \figref{fig:busch_tests}(b,c)) are significantly
more accurate, each giving an error smaller than $\sim0.01\%$. For the attractive 
branch, we
could have created more accurate pseudopotentials by decreasing the cutoff
$r_{\text{c}}$, as demonstrated in~\figref{fig:delta-cutoff}(b). Finally we
examine the bound state energy in \figref{fig:busch_tests}(d). Both the
square well and Troullier-Martins formalism give the exact ground state
energy for two atoms in a vacuum. However, the trapping potential introduces
inhomogeneity,  and the square well potential gives a $\sim10\%$ error in the
ground state energy, whereas the Troullier-Martins pseudopotential delivers 
errors of less than
$\sim0.01\%$.  This affirms the benefits of using a pseudopotential
that is robust against changes in the local environment. The success of the
Troullier-Martins and ultratransferable formalisms in describing the ground
state is all the more significant considering that these pseudopotentials aim to
describe the correct scattering properties over a range of energies. We
would therefore expect them to perform even better when modeling the excited
states of the trap.

\subsection{Excited states}

We now examine the predictions for the excited states in the repulsive and
attractive branches. Due to the shell structure, the excited states of a
few-body system are related to the ground state of a many-body
system~\cite{Bugnion13}, allowing us to probe the performance expected from
the pseudopotential in a many-body setting. We consider states up to a
maximum energy of $\Emax=7.5\hbar\omega$, corresponding to 112 non-interacting
atoms in the trap. In \figref{fig:busch_tests}(b,c) the Troullier-Martins
pseudopotential has a mean squared error averaged over all bands below
$\Emax$ between 10 and 100 times lower than the hard sphere and square well
pseudopotentials. The UTP is a further factor of 2 more
accurate. Additionally, when modeling the attractive branch, the
Troullier-Martins and ultratransferable formalisms produce
pseudopotentials that converge to the contact limit more rapidly than the
attractive square well, resulting in improved efficiency when used in a QMC
simulation.  For the cutoff radii used in~\figref{fig:busch_tests}(c),
using the Troullier-Martins pseudopotentials or the UTP
results in QMC calculations that are 4000 times more efficient than the
equivalent calculation with the square well.

The pseudopotentials deliver energies with better than ${\sim0.01}\%$ accuracy,
a significant improvement over existing pseudopotentials. This means that they 
are
no longer the limiting factor in studies of ultracold atomic gases with 
state of the art computational methods. For example, 
exact
diagonalization calculations have been performed at a similar ${\sim0.01}\%$
accuracy~\cite{Bugnion13ii}, and high fidelity many-body QMC can also achieve
${\sim0.01}\%$ stochastic error~\cite{Conduit09,Chang10,Pilati10}. 
We are therefore
well-positioned to test the pseudopotential in a many-body setting.

\section{Case study: Fermi gas}

Having demonstrated the efficacy of the Troullier-Martins and UTP formalisms
for an inhomogeneous two-body system, we now test the pseudopotentials in
quantum Monte Carlo. We calculate the equation of state of a Fermi gas with
weak interactions. Fermi gases serve as models for free electrons in a
conductor, for nucleons inside a large nucleus and for liquid
He${}^3$~\cite{Mohling61}.

For the attractive Fermi gas, the quality of a pseudopotential can be
systematically improved by reducing the cutoff radius.  We therefore
concentrate on the repulsive branch of the Feshbach resonance, for which the
top-hat pseudopotential cannot be systematically improved.  We compare the
energies predicted by DMC calculations with exact perturbation expansions
calculated with the contact potential. The main result is shown
in~\figref{fig:eq-state}: energies calculated using the UTP and top hat
differ significantly for $\kf a\gtrsim 0.3$. The equation of state
calculated using the UTP formalism agrees well with third order perturbation
theory, confirming the accuracy of the formalism.

\subsection{Formalism}

We use DMC~\cite{Ceperley80,Umrigar93,Foulkes01}, as implemented in the
\textsc{casino} code~\cite{Needs10} with a Slater-Jastrow trial wavefunction
and a backflow transformation~\cite{Rios06}. The wavefunction
takes the form $\Psi=\text{e}^J D_\uparrow D_\downarrow$, where $D_\uparrow$
and $D_\downarrow$ are Slater determinants of plane-wave orbitals for each of 
the spin
channels. The Jastrow factor $\text{e}^J$ describes interparticle
correlation,
\begin{equation*}
  J=\!\!\!\!\sum_{\substack{j\ne
      i\\\alpha,\beta\in\{\uparrow,\downarrow\}}}\!\!\!\!
  \left(1\!-\!\frac{|\pos_i\!-\!\pos_j|}{L^{\text{u}}_{\alpha\beta}}\right)^{\!\!2}\,
  u_{\alpha\beta}(|\pos_i\!-\!\pos_j|)\;\Theta(L^{\text{u}}_{\alpha\beta}-|\pos_i-\pos_j|)\punc{,}
\end{equation*}
where $u_{\alpha\beta}$ is a polynomial whose parameters we optimize in a
variational Monte Carlo (VMC) calculation, $L^{\text{u}}_{\alpha\beta}$ is a
cutoff length that we also optimize variationally and $\Theta$ is the
Heaviside step function~\cite{Drummond04}. The backflow transformation shifts 
electron positions in the Slater determinant as
\begin{equation*}
  \pos_{i\sigma} \to \pos_{i\sigma}+\!\!\!\!\sum_{\substack{j\ne
      i\\\alpha,\beta\in\{\uparrow,\downarrow\}}} \!\!\!\!
  (\pos_i-\pos_j)\,\eta^{\alpha\beta}_{ij}(|\pos_i-\pos_j|)\punc{,}
\end{equation*}
where
\begin{equation*}
  \eta^{\alpha\beta}_{ij}(r)=\left(1-\frac{r}{L^\eta_{\alpha\beta}}\right)^2
  \;\Theta(L^\eta_{\alpha\beta}-r)\; p_{\alpha\beta}(r)\punc{,}
\end{equation*}
$p_{\alpha\beta}$ is a polynomial whose parameters are optimized in
VMC, and $L^\eta_{\alpha\beta}$ is a cutoff length that we also optimize.
The backflow
transformation allows the description of further correlation, reducing the
final DMC energy~\cite{Chang10,Rios06}.

We calculate the equation of state of the Fermi gas with 81 up-spin and 81
down-spin particles. We use
twist-averaging~\cite{Rajagopal94,Rajagopal95,Lin01} and correct the
non-interacting kinetic energy with that of the corresponding infinite
system~\cite{Pilati10} to reduce finite-size effects. We use a control
variate method to reduce the stochastic error resulting from the twist
averaging procedure~\cite{Spink13}. We find that the control variate
method leads to a five-fold reduction in stochastic error for this system at
no additional computational cost.

\subsection{Results}

\begin{figure}
 \includegraphics[width=\linewidth]{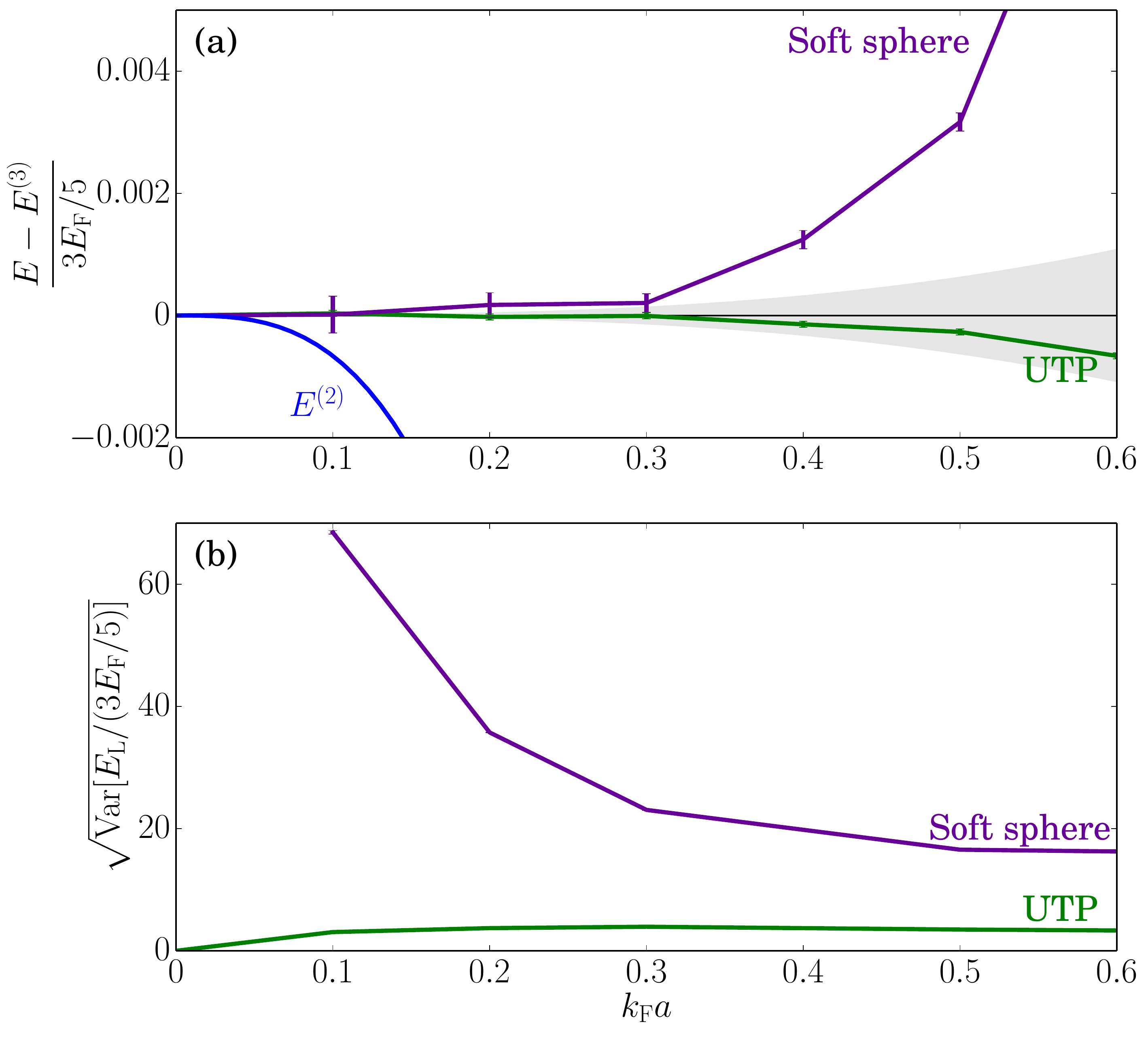}
 \caption{(Color online) (a) Deviation of the equation of state from that
   predicted by third-order perturbation theory, as given by
   Eq.~\eqref{eq:e3}. The gray region denotes the confidence intervals in
   $E^{(3)}$. The line denoted $E^{(2)}$ is the equation of state derived
   from second order perturbation theory~\cite{Huang57}. We note that the
   equation of state obtained by using a soft sphere pseudopotential
   deviates significantly from the line predicted by the perturbation expansion.
   (b) Standard deviation of the local energy $E_\mathrm{L}$ for the ground
   state of the soft sphere and the UTP. The soft sphere pseudopotential
   exhibits a much larger standard deviation, which can be explained by the
   abrupt changes in potential energy when particles overlap. }
    \label{fig:eq-state}
\end{figure}

We compare the equations of state of the UTP and soft sphere
pseudopotential in~\figref{fig:eq-state}. The two differ significantly for
$\kf a \gtrsim 0.3$, highlighting the importance of using an accurate
pseudopotential. To establish which potential reproduces the equation of
state of the contact potential more closely, we compare the results to a
third order perturbation theory calculation of the equation of
state~\cite{Mohling61,Mohling61b},
\begin{multline}
  E^{(3)} = \frac{3}{5} \Ef \Big[ 1 + \overbrace{\frac{10}{9\pi} \kf a
    + \frac{4(11 - 2\log
      2)}{21\pi^2} (\kf a)^2}^{E^{(2)}}\\
  + (0.076\!\pm\! 0.005-1/3\pi) (\kf a)^3 \Big] \punc{.}
    \label{eq:e3}
\end{multline}

\figref{fig:eq-state}(a) shows that the equation of state calculated using
the UTP remains within stochastic error of $E^{(3)}$ up to $\kf a \sim
0.6$. By contrast, the equation of state for the soft sphere system deviates
significantly from the perturbation result for $\kf a \gtrsim 0.3$. The
energy $E^{(2)}$ obtained using second order perturbation
theory~\cite{Huang57}, which is used frequently in the
literature~\cite{Duine05,Conduit08}, differs markedly from both the UTP and
soft sphere pseudopotential energy. These significant differences in energy
affirm the importance of using a pseudopotential whose scattering properties
replicate those of the contact interaction accurately.

In~\figref{fig:eq-state}(b), we compare the variance in the local energy
$E_\mathrm{L} = \hat{H}\Psi/\Psi$ of the ground state wavefunction for
different pseudopotentials. The stochastic error for a quantum Monte Carlo
calculation is proportional to $\sqrt{\mathrm{Var}(E_\mathrm{L})}$. A
smoother local energy will therefore result in more accurate results for the
same computational expense.  By virtue of its smoothness, we find that the
UTP leads to smoother local energies than the soft sphere
pseudopotential. In particular, the variance of the local energy at small
$\kf a$ diverges for the zero-range soft sphere pseudopotential.  Even at
$\kf a = 0.6$, we find that the standard deviation of the UTP ground state
is about five times smaller than that of the soft sphere ground state,
resulting in a twenty-five-fold improvement in efficiency in quantum Monte
Carlo calculations. 

The computational expense, for a fixed number of VMC or DMC samples, of all
the pseudopotentials considered in this article scales as $\mathcal{O}(N^2)$
with particle number $N$. This quadratic dependence arises from the need to
check the separation of all pairs of particles to decide whether the
particles are close enough to interact. The pre-factor of this term is
therefore identical for all pseudopotentials.  We must therefore consider
the pre-factor of the $\mathcal{O}(N)$ term to discern a difference in the
computational expense of using the pseudopotentials.  The square well or top
hat potential scales more favorably, both because it is easier to compute
the value of the pseudopotential and because, by virtue of its smaller
cutoff radius, fewer pairs of particles interact. In practice, we find that,
for the 162 particle system considered in this section at $\kf a = 0.6$, it
takes approximately 25\% less CPU time to acquire the same number of VMC
samples with a top hat potential than with a UTP. This difference is far
outweighed by the lower variance in local energy of the UTP: QMC
calculations with a top-hat pseudopotential are approximately nineteen times
more costly than calculations with a UTP, to obtain the same level of
accuracy.

\section{Discussion}

We have developed a high fidelity pseudopotential for the contact
interaction inspired by pseudopotentials common in the electronic structure
community. We tested the pseudopotential by examining the scattering phase
shifts, the energy of two trapped particles, and the ground state energy of
a Fermi gas, finding the new pseudopotentials to be approximately one
hundred times more accurate for the repulsive branch, and ten times more
accurate for the attractive and bound state branches of the Feshbach
resonance, while also 4000 times more efficient than contemporary
approximations.  The pseudopotential delivers accurate scattering properties
over all wavevectors $0\le k\lesssim\kf$ in a Fermi gas. Its smoothness also
greatly accelerates computation: for instance, for the repulsive branch of
the Feshbach resonance, calculations are accelerated by a factor of at least
nineteen.

The performance and transferability of the pseudopotential makes it widely
applicable across first principles methods including VMC, DMC, coupled
cluster and configuration interaction methods. The formalism developed can
also be applied more widely to generate pseudopotentials for narrow Feshbach
resonances, the repulsive Coulomb interaction, or the dipolar
interaction. The formalism could also be extended by using more projectors,
or the ultrasoft~\cite{Vanderbilt90} or augmented plane wave~\cite{Blochl94}
formalisms popular in the electronic structure community.

\acknowledgments{The authors thank Stefan Baur, Andrew Green, Jesper
  Levinsen, Gunnar M\"oller, Michael Rutter, and Lukas Wagner for useful
  discussions, and acknowledge the financial support of the EPSRC and
  Gonville \& Caius College.  This research used resources of the Argonne
  Leadership Computing Facility at Argonne National Laboratory, which is
  supported by the Office of Science of the U.S. Department of Energy, as
  well as resources of the Cambridge High Performance Computing Service.  }

\appendix

\section{Construction of the Troullier-Martins pseudopotential}
\label{app:TroullierMartins}

The equations in the Troullier-Martins paper pertain to electron-ion
pseudopotentials in the context of the Born-Oppenheimer
approximation~\cite{Troullier91}. They therefore consider the interaction of
an electron with a much heavier nucleus. By contrast, in this paper, we are
interested in the interaction between two particles of equal mass. This
corresponds to using a reduced mass $\mu= 1/2$ in the center of mass frame,
rather than $\mu\simeq 1$ for electron-ion pseudopotentials. We adapt the
Troullier-Martins formalism to the construction of a pseudopotential for two
particles of equal mass interacting with a contact interaction.

The Schr\"odinger equation for relative motion with reduced mass of $1/2$
and a spherically symmetric inter-particle potential is
\begin{equation*}
 [-\nabla^2 + V(\pos)]\psi_{E,\ell}(\pos)=E \psi_{E,\ell}(\pos)\punc{,}
\end{equation*}
where $\pos=\pos_1-\pos_2$ is the relative position of the two particles and
$\psi_{E,\ell}$ is the relative wavefunction associated with energy
eigenvalue $E$ and angular momentum channel $\ell$. We only consider
particles with relative angular momentum quantum number $\ell = 0$, since
the contact interaction only scatters in this channel.

By expanding the relative wavefunction $\psi_{E,\ell=0}(\pos) =
R_{E,\ell=0}(r)\mathrm{Y}_0$, where $\mathrm{Y}_0\!=\! 1/\sqrt{4\pi}$ is the
zero-th spherical harmonic and $r\!=\!|\pos|$, we can re-cast the
three-dimensional Schr\"odinger equation as a radial equation,
\begin{equation}
    \label{eq:red-se}
    \left[-\frac{1}{r^2}\pdiff{}{r}\!\left(r^2\pdiff{}{r} \right) + V(r)
    \right] R_{E,\ell=0}=E\,R_{E,\ell=0}\punc{.}
\end{equation}

We choose a calibration energy $E_{\text{c}}$, as described in the main
text, and construct the pseudopotential by choosing a
pseudo-wavefunction that matches the exact form
beyond a cutoff radius $r_{\text{c}}$, at the
calibration energy.  Following Troullier-Martins, we define the pseudo-wavefunction at the
calibration energy as $\psi^\mathrm{PP}_{E_{\text{c}},\ell=0} =
R^\mathrm{PP}_{E_{\text{c}},\ell=0} \mathrm{Y}_0$ with radial component,
\begin{equation*}
    R_{E_{\text{c}},\ell=0}^\mathrm{PP}(r) = \left\{ \begin{array}{ll}
        R_{E_{\text{c}},\ell=0}^\mathrm{cont}(r) & r \ge r_{\text{c}}   \\
        \exp[p(r)] & r < r_{\text{c}}
    \end{array} \right.
    \punc{,}
\end{equation*}
where $p(r) = \sum_{i=0}^6 c_i r^{2i} $ is a polynomial and
$R_{E_{\text{c}},\ell=0}^\mathrm{cont}$ is the radial wavefunction for the
contact interaction at the calibration energy $E_{\text{c}}$. Inserting this
form into the radial equation, Eq.~\eqref{eq:red-se}, we calculate an
expression for the pseudopotential $V^\mathrm{PP}$ as a function of $p(r)$,
\begin{equation}
    V^\mathrm{PP}(r) = \left\{ \begin{array}{ll}
            0 & r \ge r_{\text{c}} \\
            E_{\text{c}} + p'' + p'^2 + \frac{2}{r} p' & r < r_{\text{c}} 
        \end{array} \right.
    \punc{,}
    \label{eq:pp}
\end{equation}
where the primes indicate derivatives.

To proceed further, we must consider an explicit functional form for
$R^\mathrm{cont}_{E_{\text{c}},\ell=0}$. This depends on whether we are
constructing a pseudopotential for a scattering state or the bound state of
the contact interaction.  We consider these two cases in sections
\ref{sec:scattering} and \ref{sec:bound}.

\subsection{Scattering states}
\label{sec:scattering}

The relative wavefunction for two particles interacting via contact
interactions is,
\begin{equation*}
    R^\mathrm{cont}_{E,\ell=0}(r) = \frac{\sin[kr + \delta_0(k)]}{kr}
    \punc{,}
\end{equation*}
where $\delta_0(k)=\arctan(-ka)$ and $k = \sqrt{E}$. The continuity
equations at the cutoff are,
\begin{align*}
    p(r_{\text{c}}) =& \; \log\left\{ \frac{\sin[kr_{\text{c}}+\delta_0(k)]}{r_{\text{c}}} \right\}\\
    p'(r_{\text{c}}) =& \; \frac{k}{\tan(kr_{\text{c}}+\delta)} - \frac{1}{r_{\text{c}}} \\
    p''(r_{\text{c}}) =& \; -k^2 -\frac{2}{r_{\text{c}}} p' - p'^2 \\
    p^{(3)}(r_{\text{c}}) =&\;  \frac{2}{r_{\text{c}}^2} p' -
    \frac{2}{r_{\text{c}}}p'' -2p'p'' \\
    p^{(4)}(r_{\text{c}}) =&\;  - \frac{4}{r_{\text{c}}^3} p' +
    \frac{4}{r_{\text{c}}^2} p'' - \frac{2}{r_{\text{c}}} p^{(3)} -
    2p''^2-2p'p^{(3)} \punc{,}
\end{align*}
where $p^{(i)}$ denotes the $i$-th derivative of $p$ and all derivatives are
evaluated at $r = r_{\text{c}}$. To obtain the pseudo-wavefunction at the
calibration energy, we solve this system of five equations, as well as the
norm-conservation condition and impose $c_2^2 = -5c_4$ to guarantee
$\partial^2V^\mathrm{PP}/\partial r^2|_{r=0}=0$. This uniquely determines
the polynomial $p(r)$, which, in turn, determines the pseudopotential,
following Eq.~\eqref{eq:pp}.

\subsection{Bound state}
\label{sec:bound}

The relative wavefunction for two particles in the bound state of the
contact interaction is,
\begin{equation*}
    R^\mathrm{cont}_{E,\ell=0}(r) = \left(\frac{k^3}{2\pi} \right)^{1/2}
    \frac{\exp(kr)}{kr}\punc{,}
\end{equation*}
where $k=\sqrt{E}$ and $E=-1/a^2$ for scattering length $a$. The continuity
equations at the cutoff $r=r_{\text{c}}$ are
\begin{align*}
    p(r_{\text{c}}) =& \; -\frac{k}{r_{\text{c}}} - \log(r_{\text{c}}) \\
    p'(r_{\text{c}}) =& \; -k - \frac{1}{r_{\text{c}}} \\
    p''(r_{\text{c}}) =& \; -k^2 -\frac{2}{r_{\text{c}}} p' - p'^2 \\
    p^{(3)}(r_{\text{c}}) =&\;  \frac{2}{r_{\text{c}}^2} p' -
    \frac{2}{r_{\text{c}}}p'' -2p'p'' \\
    p^{(4)}(r_{\text{c}}) =&\;  - \frac{4}{r_{\text{c}}^3} p' +
    \frac{4}{r_{\text{c}}^2} p'' - \frac{2}{r_{\text{c}}} p^{(3)} -
    2p''^2-2p'p^{(3)} \punc{.}
\end{align*}


\begin{thebibliography}{99}

\bibitem{Astrakharchik04i}
G.E.~Astrakharchik, J.~Boronat, J.~Casulleras, and S.~Giorgini,
Phys. Rev. Lett. {\bf 93}, 200404 (2004).

\bibitem{Astrakharchik04ii}
G.E.~Astrakharchik, D.~Blume, S.~Giorgini, and B.E.~Granger,
Phys. Rev. Lett. {\bf 92}, 030402 (2004).

\bibitem{Conduit09}
G.J.~Conduit, A.G.~Green, and B.D.~Simons,
Phys. Rev. Lett. {\bf 103}, 207201 (2009).

\bibitem{Pilati10}
S.~Pilati, G.~Bertaina, S.~Giorgini, and M.~Troyer,
Phys. Rev. Lett. {\bf 105}, 030405 (2010).

\bibitem{Chang10}
S.-Y.~Chang, M.~Randeria, and N.~Trivedi,
Proc. Natl. Acad. Sci. USA {\bf 108}, 51 (2011).

\bibitem{Giorgini99}
S.~Giorgini, J.~Boronat, and J.~Casulleras,
Phys. Rev. A {\bf 60}, 5129 (1999).

\bibitem{Chin09}
C.~Chin, R.~Grimm, P.~Julienne, and E.~Tiesinga,
Rev. Mod. Phys. {\bf 82}, 1225 (2009).

\bibitem{Morris10}
 A.J.~Morris, P.~L\'opez R\'ios, and R.J.~Needs, Phys. Rev. A {\bf 81}, 033619 (2010).

\bibitem{Bugnion13} P.O.~Bugnion, J.A.~Lofthouse, and G.J.~Conduit,
Phys. Rev. Lett. {\bf 111}, 045301 (2013).

\bibitem{Bugnion13i} P.O.~Bugnion and G.J.~Conduit,
Phys. Rev. A {\bf 88}, 013601 (2013).

\bibitem{Bugnion13ii} P.O.~Bugnion and G.J.~Conduit,
Phys. Rev. A {\bf 87}, 060502(R) (2013).

\bibitem{Conduit09i}
G.J.~Conduit and B.D.~Simons,
Phys. Rev. Lett. {\bf 103}, 200403 (2009).

\bibitem{Jo09}
 G.-B.~Jo \emph{et al.}, Science {\bf 325}, 1521 (2009).

\bibitem{Ceperley80}
D.M.~Ceperley and B.J.~Alder,
Phys. Rev. Lett. {\bf 45}, 566 (1980).

\bibitem{Umrigar93}
    C.J.~Umrigar, M.P.~Nightingale, and K.J.~Runge, J. Chem. Phys. {\bf 99}, 2865
    (1993).

\bibitem{Foulkes01}
    W.M.C.~Foulkes, L.~Mitas, R.J.~Needs, and G.~Rajagopal, Rev. Mod. Phys.
    {\bf 73}, 33 (2001).

\bibitem{Symmetry_Constraints_DMC} W.M.C.~Foulkes, R.Q.~Hood, and R.J.~Needs,
Phys. Rev. B \textbf{60}, 4558 (1999).

\bibitem{Conduit08}
 G.J.~Conduit and B.D.~Simons, Phys. Rev. A {\bf 79}, 053606 (2009).

\bibitem{Zhai09}
H.~Zhai, Phys. Rev. A {\bf 80}, 051605(R) (2009).

\bibitem{Maslov09}
    D.L.~Maslov and A.V.~Chubukov, Phys. Rev. B {\bf 79}, 075112 (2009).

\bibitem{Chubukov09}
    A.V.~Chubukov and D.L.~Maslov, Phys. Rev. Lett. {\bf 103}, 216401 (2009).

\bibitem{Pedder13}
    C.J.~Pedder, F.~Kr\"uger, and A.G.~Green,
    Phys. Rev. B {\bf 88}, 165109 (2013).

\bibitem{Duine05}
    R.A.~Duine and A.H.~MacDonald, Phys. Rev. Lett. {\bf 95}, 230403 (2005).

\bibitem{Chubukov10}
    D.L.~Maslov and A.V.~Chubukov, Phys. Rev. B {\bf 81}, 045110 (2010).

\bibitem{Karahasanovic12}
U.~Karahasanovic, F.~Kr\"uger, and A.G.~Green,
Phys. Rev. B {\bf 85}, 165111 (2012).

\bibitem{Balian63}
R.~Balian and N.R.~Werthamer, Phys. Rev. {\bf 131}, 1553 (1963).

\bibitem{Fay80}
D.~Fay and J.~Appel, Phys. Rev. B {\bf 22}, 3173 (1980).

\bibitem{Mathur98}
N.D.~Mathur {\it et~al.}, Nature {\bf 394}, 39 (1998).

\bibitem{Roussev01}
R.~Roussev and A.J.~Millis, Phys. Rev. B {\bf 63}, 140504(R) (2001).



\bibitem{Conduit13}
G.J.~Conduit, C.J.~Pedder, and A.G.~Green,
Phys. Rev. B {\bf 87}, 121112(R) (2013).

\bibitem{Saxena00}
S.S.~Saxena, 
P.~Agarwal, K.~Ahilan, F.M.~Grosche,
R.K.W.~Haselwimmer, M.J.~Steiner, E.~Pugh, I.R.~Walker, S.R.~Julian, P.~Monthoux, 
G.G.~Lonzarich, A.~Huxley, I.~Sheikin, D.~Braithwaite, and J.~Flouquet,
Nature (London) {\bf 406}, 587 (2000).

\bibitem{Huxley01}
A.~Huxley \emph{et al.}, Phys. Rev. B {\bf 63}, 144519 (2001).

\bibitem{Hamann79}
    D.R.~Hamann, M.~Schl\"uter, and C.~Chiang, Phys. Rev. Lett. {\bf 43},
    1494 (1979).

\bibitem{Bachelet82}
    G.B.~Bachelet, D.R.~Hamann, and M.~Schl\"uter, Phys. Rev. B {\bf 26},
    4199 (1982).

\bibitem{Hamann89}
    D.R.~Hamann, Phys. Rev. B {\bf 40}, 2980 (1989).







\bibitem{Busch98}
    T.~Busch, B.G.~Englert, K.~Rzazewski, and M.~Wilkens,
    Foundations of Physics {\bf 28}, 549 (1998).

\bibitem{Troullier91}
 N.~Troullier and J.L.~Martins, Phys. Rev. B {\bf 43}, 1993 (1991).

\bibitem{Zunger79}
    A.~Zunger and M.L.~Cohen, Phys. Rev. B {\bf 20}, 4082 (1979).



\bibitem{Rappe90}
    A.M.~Rappe, K.M.~Rabe, E.~Kaxiras, and J.D.~Joannopoulos, Phys. Rev. B {\bf
    41}, 1227 (1990).

\bibitem{Lin93}
    J.S.~Lin, A.~Qteish, M.C.~Payne and V.~Heine, Phys. Rev. B {\bf 47}, 4174
    (1993).

\bibitem{supplemental-code}
    We provide a Python program to generate the Troullier-Martins, UTP, and
    square well pseudopotentials at
    \verb=https://pypi.python.org/pypi/contactpp=.

    
\bibitem{Phillips59}
 J.C.~Phillips and L.~Kleinman, Phys. Rev. {\bf 116}, 287 (1959).

\bibitem{Kleinman60}
 L.~Kleinman and J.C.~Phillips, Phys. Rev. {\bf 118}, 1153 (1960).

\bibitem{Austin62}
 B.J.~Austin, V.~Heine, and L.J.~Sham, Phys. Rev. {\bf 127}, 276 (1962).
    
\bibitem{Fahy90}
    S.~Fahy, X.W.~Wang and S.G.~Louie, Phys. Rev. B {\bf 42}, 3503 (1990).

\bibitem{Vanderbilt90}
    D.~Vanderbilt, Phys. Rev. B {\bf 41}, 7892 (1990).

\bibitem{Blochl94}
    P.E.~Bl\"ochl, Phys. Rev. B {\bf 50}, 17953 (1994).

\bibitem{Serwane11}
F.~Serwane, G.~Z\"urn, T.~Lompe, T.B.~Ottenstein, A.N.~Wenz, and S.~Jochim,
Science {\bf 332}, 336 (2011).

\bibitem{Zurn12}
G.~Z\"urn, F.~Serwane, T.~Lompe, A.N.~Wenz, M.G.~Ries, J.E.~Bohn, and S.~Jochim,
Phys. Rev. Lett {\bf 108}, 075303 (2012).

\bibitem{Mohling61}
    F.~Mohling, Phys. Rev. {\bf 122}, 1062 (1961).

\bibitem{Needs10}
    R.J.~Needs, M.D.~Towler, N.D.~Drummond, and P.~L\'opez R\'ios, J. Phys.:
    Condensed Matter {\bf 22}, 023201 (2010).

\bibitem{Rios06}
    P.~L\'opez R\'ios, A.~Ma, N.D.~Drummond, M.D.~Towler, and R.J.~Needs,
    Phys. Rev. E {\bf 74}, 066701 (2006).

\bibitem{Drummond04}
    N.D.~Drummond, M.D.~Towler, and R.J.~Needs, Phys. Rev. B {\bf 70}, 235119 (2004).

\bibitem{Rajagopal94}
    G.~Rajagopal, R.J.~Needs, S.D.~Kenny, W.M.C.~Foulkes, and A.~James,
    Phys. Rev. Lett. {\bf 73}, 1959 (1994).

\bibitem{Rajagopal95}
    G.~Rajagopal, R.J.~Needs, A.~James, S.D.~Kenny, and W.M.C.~Foulkes,
    Phys. Rev. B. {\bf 51}, 10591 (1995).

\bibitem{Lin01}
    C.~Lin, F.H.~Zong, and D.M.~Ceperley,
    Phys. Rev. E {\bf 64}, 016702 (2001).

\bibitem{Spink13}
G.G.~Spink, R.J.~Needs, and N.D.~Drummond,
Phys. Rev. B {\bf 88}, 085121 (2013).







\bibitem{Mohling61b}
    F.~Mohling, Phys. Rev. {\bf 122}, 1043 (1961).

\bibitem{Huang57}
    K.~Huang and C.N.~Yang {\bf 105}, 767 (1957).

\end{thebibliography}
\end{document}